\documentclass[conference]{IEEEtran}
\IEEEoverridecommandlockouts
\usepackage{cite}
\usepackage{amsmath,amssymb,amsfonts}
\usepackage{algorithmic}
\usepackage{graphicx}
\usepackage{textcomp}
\usepackage{xcolor}
\usepackage{comment}
\usepackage{float}
\usepackage{url}
\def\BibTeX{{\rm B\kern-.05em{\sc i\kern-.025em b}\kern-.08em
    T\kern-.1667em\lower.7ex\hbox{E}\kern-.125emX}}

\makeatletter

\def\ps@IEEEtitlepagestyle{%
  \def\@oddfoot{\mycopyrightnotice}%
  \def\@evenfoot{}%
}
\def\mycopyrightnotice{%
{\footnotesize 979-8-3503-4218-5/23\$31.00©2023 European Union\hfill} 
  \gdef\mycopyrightnotice{}
}

\begin{document}
\title{Self-Supervised Super-Resolution Approach for Isotropic Reconstruction of 3D Electron Microscopy Images from Anisotropic Acquisition
{\footnotesize \textsuperscript{}}
\thanks{This work was in part supported by the Academy of Finland (\#323385), the Erkko Foundation, and the Doctoral Programme in Molecular Medicine at the University of Eastern Finland.}
}
\author{
    \IEEEauthorblockN{Mohammad Khateri, Morteza Ghahremani, Alejandra Sierra, and Jussi Tohka}
    \IEEEauthorblockA{A. I. Virtanen Institute for Molecular Sciences, University of Eastern Finland, Finland
    \\\{mohammad.khateri, morteza.ghahremani, alejandra.sierralopez, jussi.tohka\}@uef.fi}
}

\maketitle
\begin{abstract}
Three-dimensional electron microscopy (3DEM) is an essential technique to investigate volumetric tissue ultra-structure. Due to technical limitations and high imaging costs, samples are often imaged anisotropically, where resolution in the axial direction ($z$) is lower than in the lateral directions $(x,y)$. This anisotropy 3DEM can hamper subsequent analysis and visualization tasks. To overcome this limitation, we propose a novel deep-learning (DL)-based self-supervised super-resolution approach that computationally reconstructs isotropic 3DEM from the anisotropic acquisition. The proposed DL-based framework is built upon the U-shape architecture incorporating vision-transformer (ViT) blocks, enabling high-capability learning of local and global multi-scale image dependencies. To train the tailored network, we employ a self-supervised approach. Specifically, we generate pairs of anisotropic and isotropic training datasets from the given anisotropic 3DEM data. By feeding the given anisotropic 3DEM dataset in the trained network through our proposed framework, the isotropic 3DEM is obtained. Importantly, this isotropic reconstruction approach relies solely on the given anisotropic 3DEM dataset and does not require pairs of co-registered anisotropic and isotropic 3DEM training datasets. To evaluate the effectiveness of the proposed method, we conducted experiments using three 3DEM datasets acquired from brain. The experimental results demonstrated that our proposed framework could successfully reconstruct isotropic 3DEM from the anisotropic acquisition.
\end{abstract}


\section{Introduction}
Three-dimensional electron microscopy (3DEM) enables the visualization and analysis of volumetric tissue ultrastructure at nanometer resolution. Achieving isotropic acquisition, where resolution is consistent in all dimensions, can assist downstream image analysis and visualization tasks. However, practical limitations, such as the constraints of EM techniques and imaging time and costs, often lead to achieving the resolution in the axial $(z)$  direction lower than lateral $(x,y)$ directions. Focused ion beam scanning EM (FIB-SEM) is one EM technique that can obtain isotropic 3DEM images with sub-10nm resolution in all directions; however, FIB-SEM is low-throughput. On the other hand, serial section transmission EM (ssTEM) or serial block-face scanning EM (SBEM) offers higher throughput and cost-effectiveness compared to FIB-SEM but cannot achieve the required axial resolution ~\cite{mikula2016progress}. Image super-resolution (SR) is a computational approach that can increase the axial resolution to match lateral resolutions, enabling the reconstruction of isotropic 3DEM from anisotropic acquisitions. 

Traditional SR approaches rely on interpolation methods, which can increase axial resolution. However, these methods have limitations in recovering fine missing details in low-resolution (LR) axial planes $(xz/yz)$. To overcome these limitations, learning-based methods have been proposed that leverage prior knowledge about the latent data to the interpolation. One such method is sparse representation over learned dictionaries, which has been used in various SR applications \cite{song2019multimodal, yang2010image}. However, since dictionaries are learned from small image patches, they may not reconstruct high-quality EM images with large field-of-view. Authors in ~\cite{hu2012super} proposed a dictionary-learning-based approach to reconstruct isotropic 3DEM by combining anisotropic 3DEM with sparse tomographic views of the same sample acquired at a finer axial resolution. While this approach offered a promising solution for isotropic reconstruction of 3DEM, it relies on the availability of both anisotropic and sparse tomographic views, which may not always be feasible.
\begin{figure*}
\centering
\includegraphics[width=0.85\textwidth]{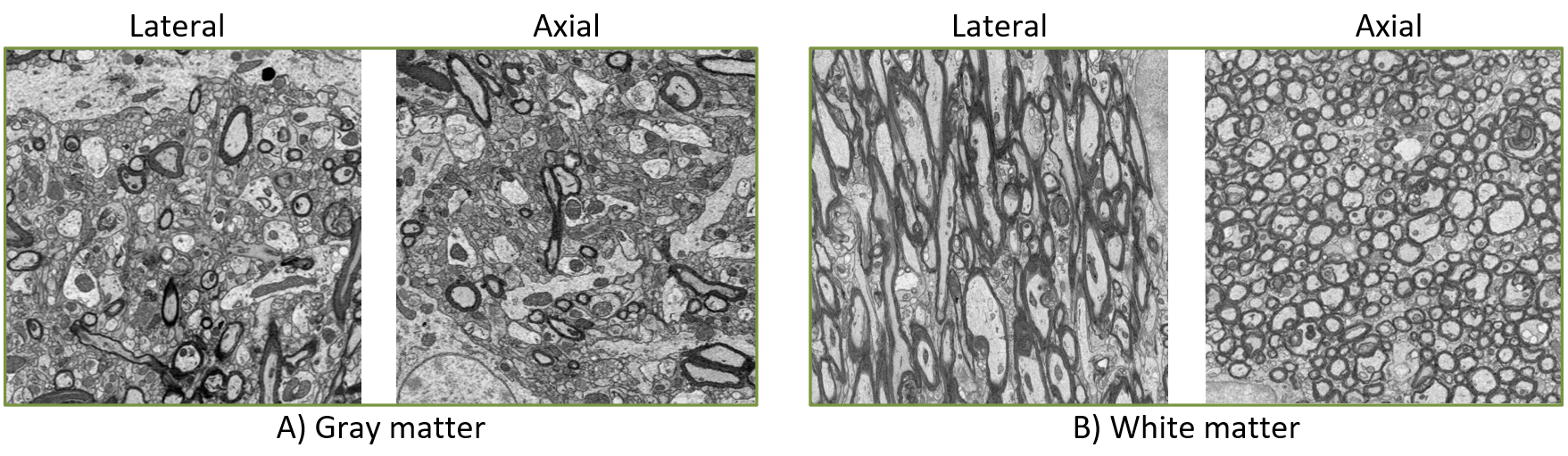}

\caption{Ultra-structural self-similarity in 3DEM datasets from the rat brain. A) Gray matter \cite{salo2018quantification} demonstrates the ultra-structural self-similarity across a wide range of sizes, while B) white matter \cite{abdollahzadeh2021deepacson} predominantly exhibits this self-similarity in smaller structures.} \label{fig_selfsimilarity}
\end{figure*}

Deep learning (DL) has emerged as a promising approach for SR in computer vision ~\cite{chen2022real}, medical ~\cite{sui2022scan}, and biomedical ~\cite{fang2021deep} applications. DL-based methods follow an end-to-end learning procedure, enabling them to effectively learn the mappings from LR to high-resolution (HR) spaces when abundant LR and HR training datasets are available. The DL-based approach for isotropic 3DEM reconstruction from the anisotropic acquisition was introduced in ~\cite{heinrich2017deep}, in which authors adopted a 3D convolutional neural network (CNN) architectures, then trained it using pairs of down-sampled isotropic 3DEM (synthetic anisotropic) and isotropic 3DEM acquired from FIB-SEM and tested on images obtained from the same technology. However, this approach has some limitations. Importantly, it requires the availability of isotropic 3DEM images at the desired resolution, which is often not feasible -- especially in ssTEM and SBEM techniques. Additionally, when the network is fed with anisotropic 3DEM images acquired from a different technology, severe performance drops, and artifacts may occur due to the domain gap between EM imaging techniques. 

Self-supervised super-resolution learning is a powerful technique that can eliminate the need for training datasets and address the domain gap between training and test datasets. It involves training super-resolution algorithms solely on the given LR image, using synthetically generated LR-HR training pairs derived from the LR image itself. Authors in ~\cite{shocher2018zero} introduced the concept of self-supervised super-resolution learning, where they harnessed the internal recurrence of information inside a given LR natural image across different resolution scales to generate synthetic pairs of LR and HR image datasets. When the network is trained, the given LR image is fed to the network to produce the corresponding HR image. This approach has been employed within studies in the biomedical~\cite{ weigert2017isotropic} and medical~\cite{zhao2020smore} domains to produce 3D isotropic images from the anisotropic acquisition, respectively, with the focus on the optical fluorescence microscopy and magnetic resonance imaging.

Motivated by the remarkable self-similarity observed in ultra-structures within brain 3DEM datasets, we present an efficient self-supervised super-resolution framework specifically designed to transform anisotropic 3DEM data into isotropic 3DEM, named A2I-3DEM. The key contributions of our work are as follows:

\begin{itemize}
    \item We propose a framework for reconstructing isotropic 3DEM data from anisotropic acquisition while mitigating the inherent noise-like artifacts present in electron microscopy.
        
    \item We introduce a novel DL architecture based on the vision transformer, which effectively captures multi-scale local and global image dependencies, helping in enhanced reconstruction. 

    \item We employ an efficient training strategy by simulating the distortions commonly observed in 3DEM imaging.
\end{itemize}

\section{Method}
Let $\boldsymbol{x} \in \mathbb{R}^{W \times W \times W}$ and $\boldsymbol{y} \in \mathbb{R}^{W \times W \times C}$ denote respectively isotropic and anisotropic 3DEM, where $\rho = W/ C$ indicates the resolution ratio between isotropic and anisotropic acquisitions in the axial direction ($z$), i.e., super-resolution ratio. In this section, we introduce our ViT-empowered self-supervised super-resolution approach to reconstruct isotropic 3DEM, $\boldsymbol{x}$, from the anisotropic acquisition $\boldsymbol{y}$.

\begin{figure*}[ht]
\centering
\includegraphics[width=0.73\textwidth]{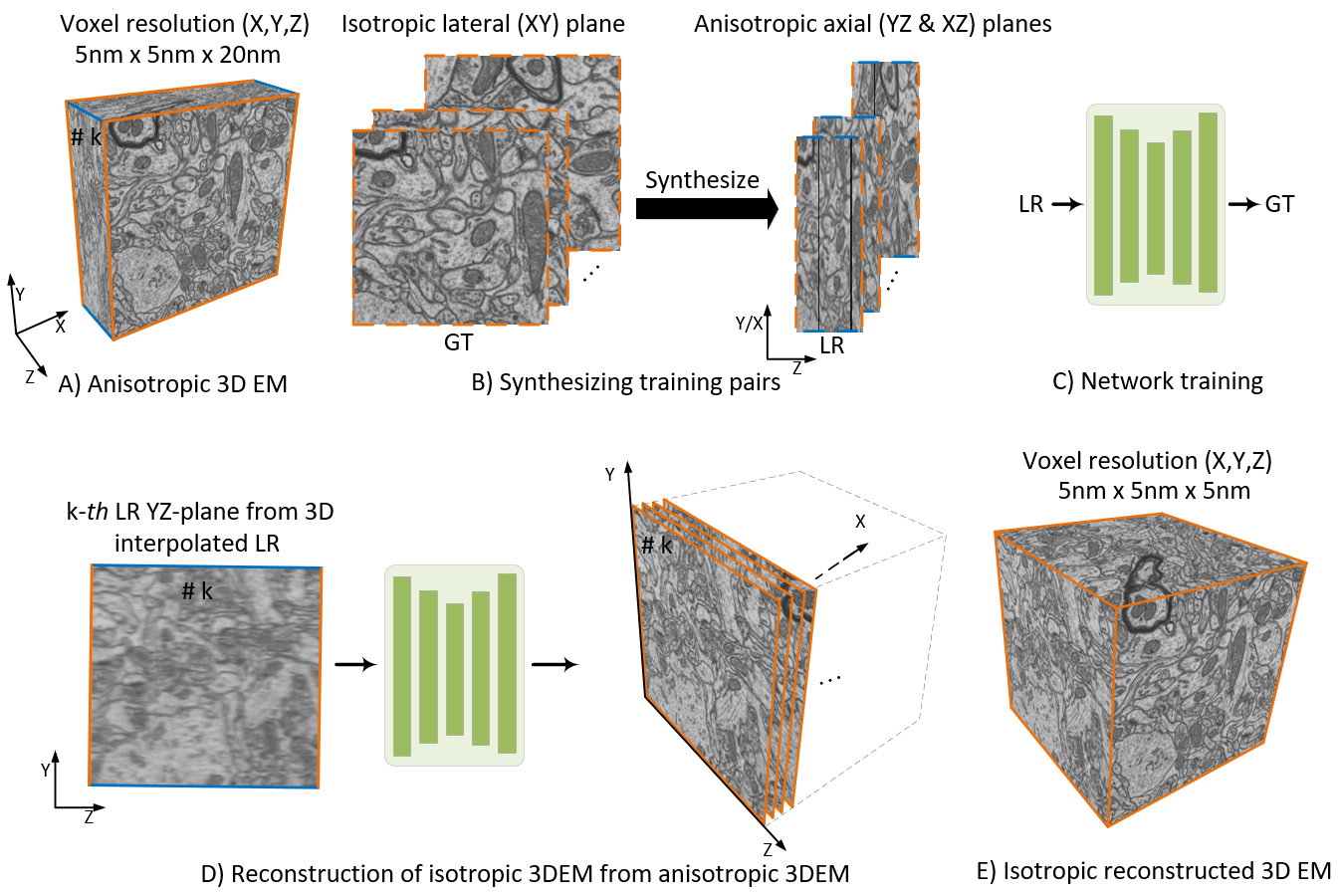}
\caption{ The workflow of the proposed self-supervised super-resolution framework for the isotropic reconstruction of 3DEM from the anisotropic acquisition. A) Input 3DEM data is anisotropic, with high resolution in the lateral $(x,y)$ directions and inferior resolution in the axial direction ($z$). B) Training pairs are synthesized from the anisotropic 3DEM data. The isotropic $xy-$lateral plane  undergoes under-sampling and distortions to synthesize the $xz-$ and $yz-$anisotropic  axial planes. C) The proposed network is trained using synthesized training pairs, where the interpolated synthesized axial plane is employed as LR input, while the isotropic lateral plane is regarded as GT. D) The trained network sequentially takes each axial plane as input, and the resultant outputs are stacked together to obtain isotropic 3DEM, involving two steps: Initially, 3D interpolation is employed to resize the anisotropic 3DEM data, aligning it with the size of the desire 3D isotropic data. Subsequently, the trained network is consecutively fed with each slice from the interpolated data's axial plane, and the resultant outputs are stacked together to generate isotropic 3DEM with an improved resolution in the axial direction. E) The output is an isotropic 3DEM with the improved resolution ratio $\rho$ in the axial direction.}
\label{fig1}
\end{figure*}

\begin{figure*}[ht]
\centering
\includegraphics[width=0.59\textwidth]{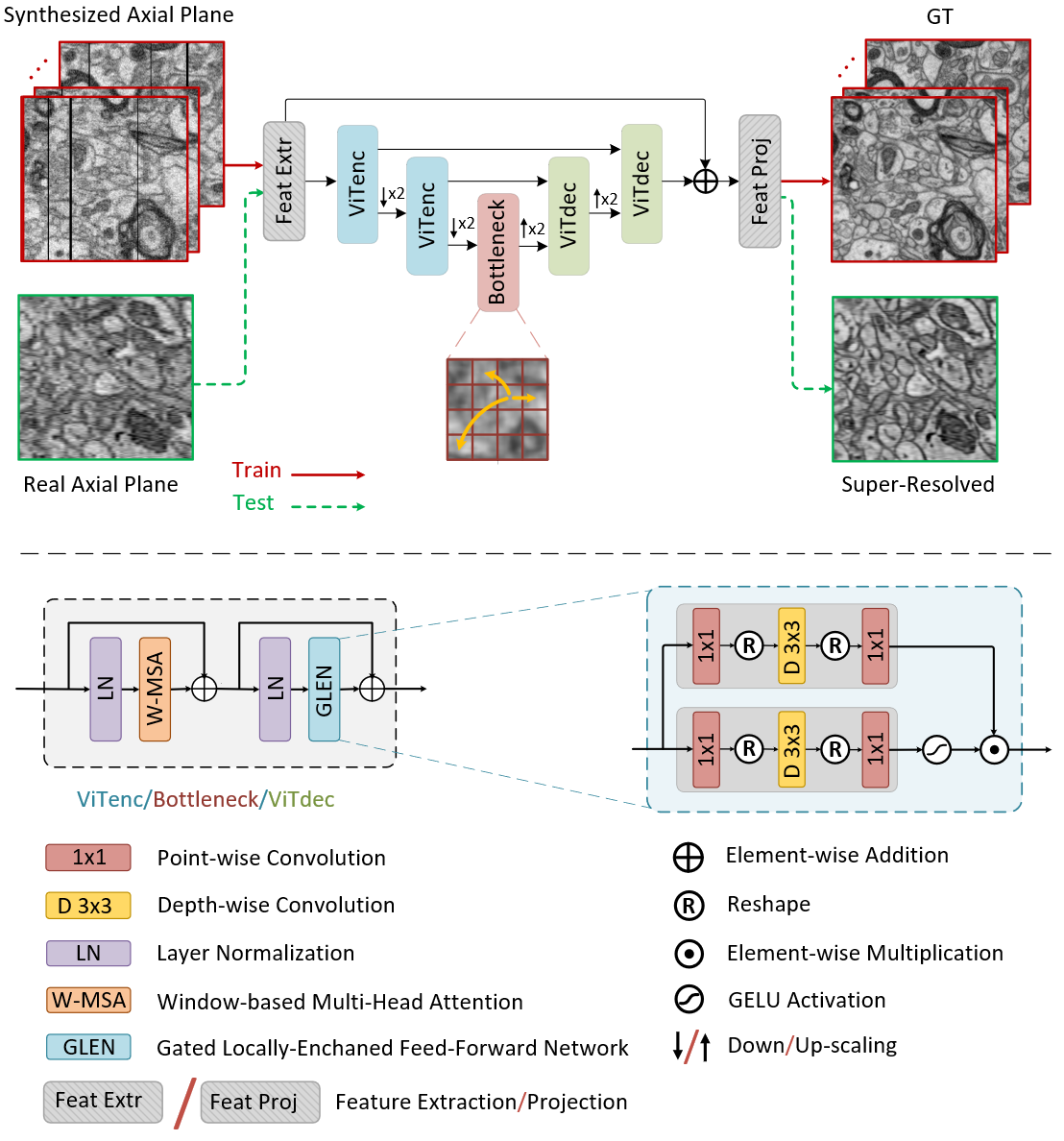}
\caption{
The proposed U-shaped architecture based on the vision transformer. Training and testing are illustrated in the upper part of the figure, marked respectively in red and green. The bottom part of the figure visualizes the component of the proposed architecture. 
} \label{fig_3}
\end{figure*}

\subsection{Self-Supervised Super-Resolution}
Self-similarity of ultra-structures between lateral and axial planes in 3DEM data, especially in the brain gray matter, allows for self-supervised learning upon the anisotropic 3DEM data, see Fig \ref{fig_selfsimilarity}. Leveraging such a structural self-similarity, we can synthesize training image pairs from the isotropic $xy-$lateral plane. To synthesize the training pairs, large patches that adequately represent the ultrastructural features of interest are extracted from the lateral plane ${P_{xy}^{i} \in \mathbb{R}^{M \times M}}$. These patches are then subjected to various degradations such as noise, artifacts, distortions, and anisotropic under-sampling resolution with ratio $1/\rho$ to generate corresponding synthesized axial patches ${P_{xz/yz}^{i} \in \mathbb{R}^{(M/\rho) \times M}}$. The synthesized pairs $\{(P_{xz/yz}^{i},P_{xy}^{i})\}_{i=1}^{N}$ are then used to train network $f_{\theta}(\cdot): \mathbb{R}^{(M/\rho) \times M} \to \mathbb{R}^{M \times M}$ parameterized with $\theta$ to learn the mapping from axial to lateral planes. In practice, $\rho$ may not always be an integer, which poses challenges to determining the mapping. To overcome this issue, we first employ interpolation to resize the anisotropic data to match the desired isotropic data size. This interpolated data is then utilized as the LR image. The network's parameters $\theta$ are obtained by optimizing the following empirical loss:
\begin{equation}
\hat{\theta} =\arg\min_{\theta} \sum_{i=1}^{N} \mathcal{L}(f_{\theta}(P_{xz/yz}^i),P_{xy}^i), 
\end{equation}
where $\mathcal{L}$ is the loss function between network prediction $f_{\theta}(P_{xz/yz})$ and ground truth $P_{xy}$. The trained network $f_{\theta}(\cdot)$ is then used to super-resolve the real axial planes to the desired resolution. Finally, by stacking the super-resolved axial planes in the perpendicular direction, the isotropic 3DEM is reconstructed. The proposed self-supervised super-resolution framework is illustrated in  Fig.\ref{fig1}.

\subsection{Network Architecture}
\subsubsection{Overall Pipeline} The proposed network architecture is a hierarchical U-shaped design of the encoder-decoder equipped with ViT blocks, as illustrated in Fig.\ref{fig_3}. The input is a low-resolution axial plane image, $\bold{I} \in \mathbb{R}^{1 \times H \times W}$, which is first fed through convolutional layers to extract low-level features, $\bold{X_0} \in \mathbb{R}^{C \times H \times W}$, where $C$, $H$, and $W$ respectively indicate the number of channels, height, and width. Afterward, the feature map is passed through a symmetric encoder-decoder with $K$ levels. Starting from the first encoder, the encoder hierarchically reduces the spatial resolution ($H \times W$) while increasing the channel size, leading to the bottleneck feature map, $\bold{F_{\ell}} \in \mathbb{R}^{2^{K-1} C \times \frac{H}{2^{K-1}} \times \frac{W}{2^{K-1}}}$. The feature maps from the bottleneck and encoders are then passed to the decoders to progressively produce the high-resolution representation. Finally, the low-level features are added to the output from the last decoder, and fed with to the feature projection block, producing the super-resolved image.

\subsubsection{Vision Transformer}
The ViTs partition an image into a sequence of small patches, i.e., local windows, and learn relationships between them. By learning these relationships, the ViT can learn a wide range of image dependencies, which is crucial for achieving high performance in low-level vision tasks like image super-resolution. To capture both global and local image dependencies while keeping computational costs low, we employ the window-based multi-head attention (W-MSA) approach ~\cite{wang2022uformer, liang2021swinir}. The extracted attention maps using W-MSA are then passed through the novel gating mechanism, called the gated locally-enhanced feed-forward network (GLEN), to  enhance the important features while suppressing the less important ones. These W-MSA and GLEN are embedded into a ViT block illustrated in Fig.\ref{fig_3}, and the corresponding computation is as follows:
\begin{equation}
\begin{split}
&\bold{X^{\prime}} = \textbf{W-MSA}(\bold{LN}(\bold{X})),\\
&\bold{X^{\prime\prime}} =  \textbf{GLEN}(\bold{LN}(\bold{X}^{\prime}))+ \bold{X}^{\prime},
\end{split}
\end{equation}
where, $\bold{LN}$ is layer normalization  and $\bold{X}$ is the input feature map. 

\paragraph{W-MSA} The input feature map $\bold{X} \in \mathbb{R}^{C \times H \times W} $ is firstly partitioned into $N=HW/M^2$ non-overlapping $M \times M$ local windows, leading to the local feature map $\bold{X}^{i} \in \mathbb{R}^{M^2 \times C}$. The standard self-attention mechanism is then applied to each local feature map. The W-MSA, when there is $k$ head with the dimension of $d_k=C/k$, is obtained by concatenating attention heads $\bold{\hat{X}}_{k} = \{\bold {Y}^{i}_{k}\}_{i=1}^{N}$, where $\bold{Y}^{i}_{k}$ is $k$-th head attention related to $i$-th local window calculated as below:
\begin{equation}
\bold{Y}_{k}^{i} = \textbf{Attention} ( \bold {X}^{i} \bold {W}_{k}^{Q}, \bold {X}^{i} \bold {W}_{k}^{K}, \bold {X}^{i} \bold {W}_{k}^{V}), i=1,\dots, N,
\end{equation}
where $\bold{W}_{k}^{Q}$, $\bold{W}_{k}^{K}$, $\bold{W}_{k}^{V} \in \mathbb{R}^{C \times d_k} $ are projection metrices of queries ($\bold{Q}$), keys ($\bold{K}$), and values ($\bold{V}$) for the $k$-th head, respectively. The attention is obtained as follows:
\begin{equation}
\textbf{Attention}(\bold{Q},\bold{K},\bold{V}) = \textbf{SoftMax}(\frac{\bold{Q}\bold{K}^{T}}{{\sqrt{d_k}}}+ \bold{B})\bold{V},
\end{equation}
where $\bold{B}$ is the relative position bias \cite{liu2021swin}.

\paragraph{GLEN} This block processes attention maps through two components: depth-wise convolution, which learns contextual image dependencies required for SR, and a gating mechanism, which highlights informative features while suppressing non-informative ones. As shown in Fig.\ref{fig_3}, the gating mechanism is implemented as the element-wise product of two parallel paths of linear transformation layers.

\subsubsection{Loss Function} To optimize the network's parameters, we utilize the $\mathcal{L}_{\ell_1} = \frac{1}{N} \sum_{i=1}^{N}{{\Vert x_i - \hat{x_i} \Vert}_1}$ and projected distribution loss (PDL) \cite{delbracio2020projected}, which respectively 
penalize pixel value and distribution mismatch between restored image $\hat{x}$ and ground truth $x$, ensuring both pixel-level accuracy and distribution-level fidelity. The total loss is given by:
\begin{equation}
\mathcal{L}_{Total} =  \mathcal{L}_{\ell1} + \alpha \mathcal{L}_{PDL} ,
\end{equation}
where $\alpha$ is a hyperparameter governing the trade-off between loss functions, which was empirically set to $0.01$. 
For optimization, we employed the Adam algorithm \cite{kingma2014adam} with an initial learning rate of $10^{-4}$. The implementation was done using PyTorch framework.

\section{Experiments and Results}

\begin{figure*}
\centering
\includegraphics[width=0.98\textwidth]{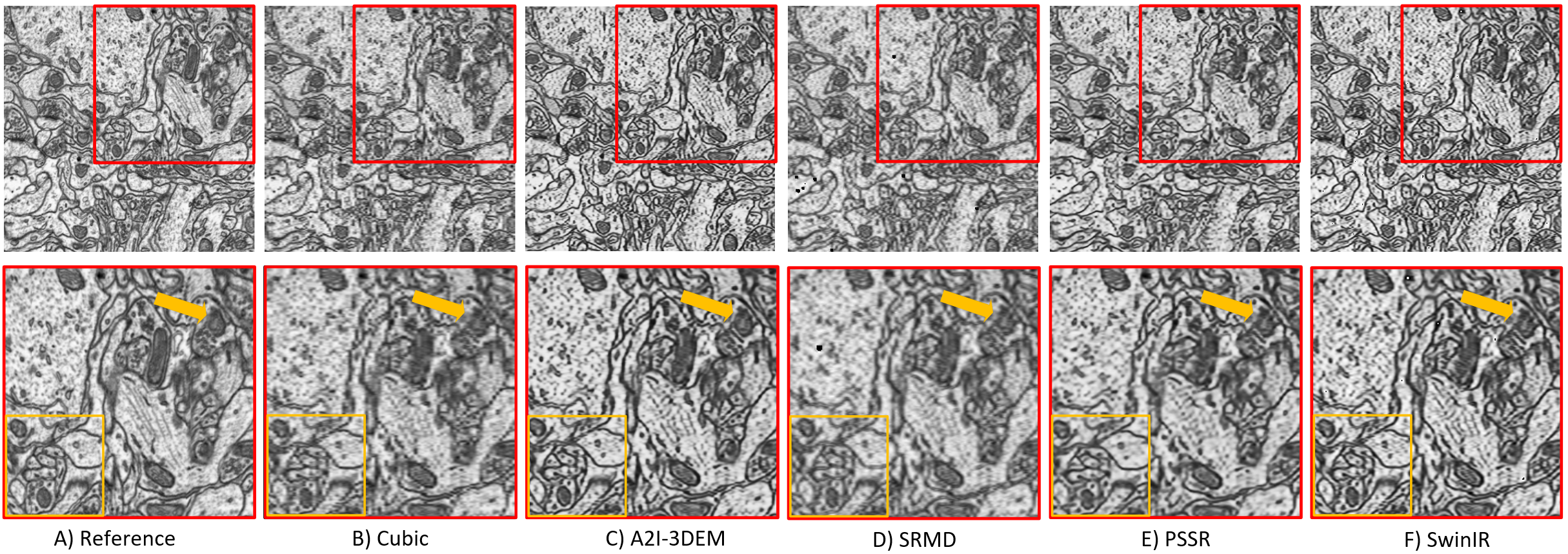}
\caption{Visual comparison of isotropic 3DEM reconstruction results using various methods on the synthetic dataset: $xz-$axial plane perspective. } \label{fig_synthetic}
\end{figure*}

\begin{figure*}
\centering
\includegraphics[width=0.98\textwidth]{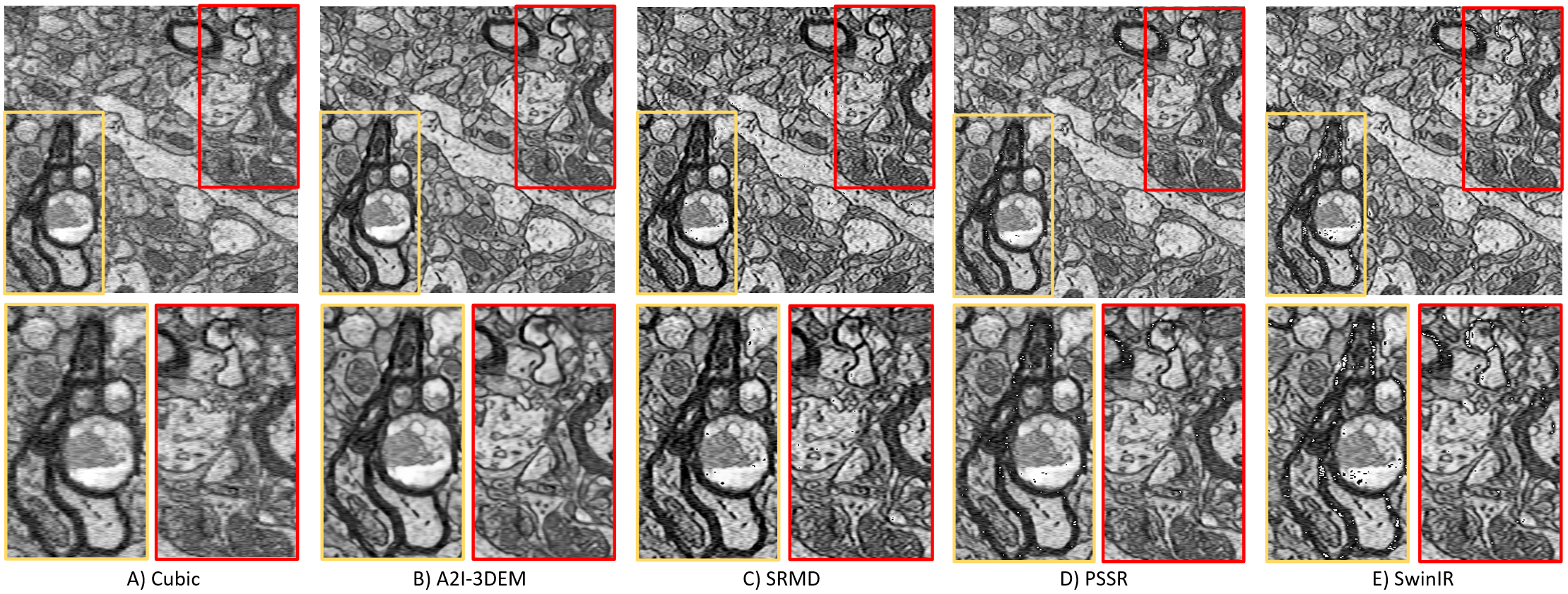}
\caption{Visual comparison of isotropic 3DEM reconstruction results  using various methods on the real dataset from gray matter: $xz-$axial plane perspective.} \label{fig_real_GM}
\end{figure*}

\subsection{Datasets}

\subsubsection{Synthetic Data} We synthesized an anisotropic 3DEM dataset by under-sampling an isotropic FIB-SEM dataset~\cite{ref_epfl}. In the first step, to reduce noise and artifacts in data, we isotropically downscaled the original data-- with voxel resolution $5 \times 5 \times 5$ $nm^3$ and image size of $1530 \times 1530 \times 1053$-- by a factor of three,  resulting in a voxel resolution of $15 \times 15 \times 15$ $nm^3$. Subsequently, we applied anisotropic downsampling to achieve a voxel resolution of $15 \times 15 \times 45$ $nm^3$. These synthetic pairs of anisotropic and isotropic 3DEM datasets were utilized in our experiments.

\subsubsection{Real Data} We used two anisotropic 3DEM datasets acquired from rat brains through the SBEM technique.
The first dataset was acquired from the gray matter in the visual cortex ~\cite{salo2018quantification} with the size of $1024 \times 1024 \times 540$, while the second was acquired from the white matter at the corpus callosum ~\cite{abdollahzadeh2021deepacson}, with the size of $1024 \times 1024 \times 490$.  Both datasets had a voxel resolution of $15 \times 15 \times 50 $ $nm^3$.

\subsection{Results}
We compared the proposed super-resolution method, A2I-3DEM, with several established techniques, including the standard cubic interpolation approach as well as two CNN-based methods: SRMD \cite{zhang2018learning} and PSSR \cite{fang2021deep}. Additionally, we considered a transformer-based method, SwinIR \cite{liang2021swinir}. For synthetic data, we utilized PSNR and SSIM ~\cite{ref_ssim} for quantitative assessments and visually compared the super-resolved volumes with the reference. For real data, lacking a reference, we visually compared the results with the Cubic interpolated data, the initial point for all competitors, to assess resolution enhancement and consistency of details.


For the synthetic dataset, where we have the reference, a visual comparison with competitors is drawn in Fig. \ref{fig_synthetic}, and corresponding quantitative results were tabulated in Table\ref{tab1}. In Fig. \ref{fig_synthetic}, orange restricted areas show that cubic and SRMD led to severely blurred results. Among other methods, A2I-3DEM and SwinIR could produce images with better contrast and distinguishable membranes. Notably, as pointed out by the arrows, A2I-3DEM outperforms SwinIR by producing outputs with reduced blurriness. The superiority of A2I-3DEM is in agreement with the PSNR value reported in Table\ref{tab1}. However, SSIM values contradict the visual outcomes, as the cubic interpolation method appears to outperform all other competitors according to SSIM. This discrepancy calls for an alternative image quality assessment metric. 

\begin{table}[ht]
\caption{Quantitative comparisons of isotropic 3DEM reconstruction on the Synthetic Dataset. The best metric value for each method is marked in bold.}\label{tab1}
\begin{center}
\begin{tabular}{| c || c | c | c | c | c | c |}
\hline
\textbf{Metric} & \multicolumn{5}{c|}{\textbf{Method}}  \\ 
\cline{2-6} & Cubic & SRMD & PSSR  & SwinIR & A2I-3DEM \\
\hline
PSNR      & 28.15 & 29.16 & 29.11  & 30.57 &  \textbf{30.61}  \\ \hline
SSIM      & \textbf{0.698} & 0.644 & 0.675  & 0.632 &  0.645    \\ \hline
\end{tabular}
\end{center}
\end{table}

Visual comparison of the first real dataset, pertaining to brain gray matter, is presented in Fig. \ref{fig_real_GM}. Consistent with expectations, DL-based methods demonstrate enhanced detail compared to cubic interpolation. Zooming in on specific regions in Fig. \ref{fig_real_GM} (B-E), artifacts such as black point artifacts in white areas or white point artifacts in black areas are evident in the results of SRMD, PSSR, and SwinIR. In contrast, A2I-3DEM not only avoids these artifacts but also successfully reduces noise compared to the other methods.

\begin{figure}
\centering
\includegraphics[width=0.49\textwidth, height=0.41\textwidth]{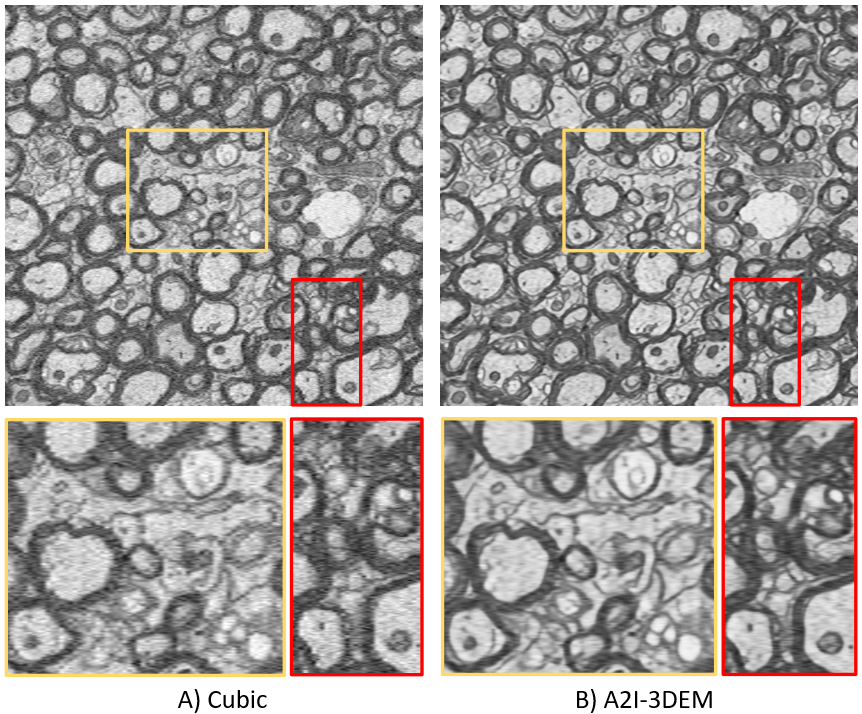}
\caption{Isotropic 3DEM reconstruction of real dataset from white matter: $xz-$axial plane perspective.} \label{fig_real_WM}
\end{figure}

A subset of visual results from the second real dataset, related to brain white matter, is depicted in Fig. \ref{fig_real_WM}. These results highlight the success of our proposed self-supervised method in enhancing the resolution of the given LR image while effectively mitigating noise.

\section{Conclusion}
This paper introduced a deep-learning-based self-supervised super-resolution framework to overcome the challenge of acquiring isotropic 3DEM. The framework's ability to generate training datasets directly from the provided anisotropic 3DEM data makes it a practical preprocessing tool for downstream visualization and processing tasks. The incorporation of simulated distortions within the efficient training strategy not only improved the model's generalizability but also enabled the network to learn to mitigate noise that exists in the given LR EM image. Furthermore, the proposed U-shaped architecture, equipped with ViT blocks, effectively captures multi-scale local and global image dependencies, leading to enhanced reconstruction performance. Experimental evaluations conducted on 3DEM datasets of brain tissue demonstrated the network's proficiency in recovering fine details while effectively mitigating noise. 

\section*{Acknowledgment}
We thank  CVLab at École Polytechnique Fédérale de Lausanne for sharing their 3DEM dataset, Electron Microscopy Unit of the Institute of Biotechnology at University of Helsinki for rat datasets,  and the Bioinformatics Center at University of Eastern Finland, for providing computational resources.

\bibliographystyle{ieeetr}
\bibliography{main.bib}

\begin{thebibliography}{10}

\bibitem{mikula2016progress}
S.~Mikula, ``Progress towards mammalian whole-brain cellular connectomics,''
  {\em Frontiers in neuroanatomy}, vol.~10, p.~62, 2016.

\bibitem{song2019multimodal}
P.~Song, X.~Deng, J.~F. Mota, N.~Deligiannis, P.~L. Dragotti, and M.~R.
  Rodrigues, ``Multimodal image super-resolution via joint sparse
  representations induced by coupled dictionaries,'' {\em IEEE Transactions on
  Computational Imaging}, vol.~6, pp.~57--72, 2019.

\bibitem{yang2010image}
J.~Yang, J.~Wright, T.~S. Huang, and Y.~Ma, ``Image super-resolution via sparse
  representation,'' {\em IEEE transactions on image processing}, vol.~19,
  no.~11, pp.~2861--2873, 2010.

\bibitem{hu2012super}
T.~Hu, J.~Nunez-Iglesias, S.~Vitaladevuni, L.~Scheffer, S.~Xu, M.~Bolorizadeh,
  H.~Hess, R.~Fetter, and D.~Chklovskii, ``Super-resolution using sparse
  representations over learned dictionaries: Reconstruction of brain structure
  using electron microscopy,'' {\em arXiv preprint arXiv:1210.0564}, 2012.

\bibitem{salo2018quantification}
R.~A. Salo, I.~Belevich, E.~Manninen, E.~Jokitalo, O.~Gr{\"o}hn, and A.~Sierra,
  ``Quantification of anisotropy and orientation in 3d electron microscopy and
  diffusion tensor imaging in injured rat brain,'' {\em Neuroimage}, vol.~172,
  pp.~404--414, 2018.

\bibitem{abdollahzadeh2021deepacson}
A.~Abdollahzadeh, I.~Belevich, E.~Jokitalo, A.~Sierra, and J.~Tohka,
  ``Deepacson automated segmentation of white matter in 3d electron
  microscopy,'' {\em Communications biology}, vol.~4, no.~1, p.~179, 2021.

\bibitem{chen2022real}
H.~Chen, X.~He, L.~Qing, Y.~Wu, C.~Ren, R.~E. Sheriff, and C.~Zhu, ``Real-world
  single image super-resolution: A brief review,'' {\em Information Fusion},
  vol.~79, pp.~124--145, 2022.

\bibitem{sui2022scan}
Y.~Sui, O.~Afacan, C.~Jaimes, A.~Gholipour, and S.~K. Warfield, ``Scan-specific
  generative neural network for mri super-resolution reconstruction,'' {\em
  IEEE Transactions on Medical Imaging}, vol.~41, no.~6, pp.~1383--1399, 2022.

\bibitem{fang2021deep}
L.~Fang, F.~Monroe, S.~W. Novak, L.~Kirk, C.~R. Schiavon, S.~B. Yu, T.~Zhang,
  M.~Wu, K.~Kastner, A.~A. Latif, {\em et~al.}, ``Deep learning-based
  point-scanning super-resolution imaging,'' {\em Nature methods}, vol.~18,
  no.~4, pp.~406--416, 2021.

\bibitem{heinrich2017deep}
L.~Heinrich, J.~A. Bogovic, and S.~Saalfeld, ``Deep learning for isotropic
  super-resolution from non-isotropic 3d electron microscopy,'' in {\em Medical
  Image Computing and Computer-Assisted Intervention- MICCAI 2017: 20th
  International Conference, Quebec City, QC, Canada, September 11-13, 2017,
  Proceedings, Part II 20}, pp.~135--143, Springer, 2017.

\bibitem{shocher2018zero}
A.~Shocher, N.~Cohen, and M.~Irani, ``“zero-shot” super-resolution using
  deep internal learning,'' in {\em Proceedings of the IEEE conference on
  computer vision and pattern recognition}, pp.~3118--3126, 2018.

\bibitem{weigert2017isotropic}
M.~Weigert, L.~Royer, F.~Jug, and G.~Myers, ``Isotropic reconstruction of 3d
  fluorescence microscopy images using convolutional neural networks,'' in {\em
  Medical Image Computing and Computer-Assisted Intervention- MICCAI 2017: 20th
  International Conference, Quebec City, QC, Canada, September 11-13, 2017,
  Proceedings, Part II 20}, pp.~126--134, Springer, 2017.

\bibitem{zhao2020smore}
C.~Zhao, B.~E. Dewey, D.~L. Pham, P.~A. Calabresi, D.~S. Reich, and J.~L.
  Prince, ``Smore: a self-supervised anti-aliasing and super-resolution
  algorithm for mri using deep learning,'' {\em IEEE transactions on medical
  imaging}, vol.~40, no.~3, pp.~805--817, 2020.

\bibitem{wang2022uformer}
Z.~Wang, X.~Cun, J.~Bao, W.~Zhou, J.~Liu, and H.~Li, ``Uformer: A general
  u-shaped transformer for image restoration,'' in {\em Proceedings of the
  IEEE/CVF Conference on Computer Vision and Pattern Recognition},
  pp.~17683--17693, 2022.

\bibitem{liang2021swinir}
J.~Liang, J.~Cao, G.~Sun, K.~Zhang, L.~Van~Gool, and R.~Timofte, ``Swinir:
  Image restoration using swin transformer,'' in {\em Proceedings of the
  IEEE/CVF international conference on computer vision}, pp.~1833--1844, 2021.

\bibitem{liu2021swin}
Z.~Liu, Y.~Lin, Y.~Cao, H.~Hu, Y.~Wei, Z.~Zhang, S.~Lin, and B.~Guo, ``Swin
  transformer: Hierarchical vision transformer using shifted windows,'' in {\em
  Proceedings of the IEEE/CVF international conference on computer vision},
  pp.~10012--10022, 2021.

\bibitem{delbracio2020projected}
M.~Delbracio, H.~Talebi, and P.~Milanfar, ``Projected distribution loss for
  image enhancement,'' {\em arXiv preprint arXiv:2012.09289}, 2020.

\bibitem{kingma2014adam}
D.~P. Kingma and J.~Ba, ``Adam: A method for stochastic optimization,'' {\em
  arXiv preprint arXiv:1412.6980}, 2014.

\bibitem{ref_epfl}
\url{https://www.epfl.ch/labs/cvlab/data/data-em/}.

\bibitem{zhang2018learning}
K.~Zhang, W.~Zuo, and L.~Zhang, ``Learning a single convolutional
  super-resolution network for multiple degradations,'' in {\em Proceedings of
  the IEEE conference on computer vision and pattern recognition},
  pp.~3262--3271, 2018.

\bibitem{ref_ssim}
Z.~Wang, A.~C. Bovik, H.~R. Sheikh, and E.~P. Simoncelli, ``Image quality
  assessment: from error visibility to structural similarity,'' {\em IEEE
  transactions on image processing}, vol.~13, no.~4, pp.~600--612, 2004.

\end{thebibliography}

\end{document}